\documentclass{aa}
\usepackage[T1]{fontenc}
\usepackage[utf8]{inputenc}
\usepackage{graphicx}
\usepackage{txfonts}
\usepackage{amsmath}
\usepackage{booktabs}
\usepackage{url}
\usepackage[hidelinks]{hyperref}

\graphicspath{{figures/}}
\newcommand{\AAincludegraphics}[2][]{%
  \IfFileExists{figures/#2}{\includegraphics[#1]{#2}}{%
    \fbox{\parbox[c][3.2cm][c]{0.89\linewidth}{\centering
    \small\sffamily [Figure not provided]\\[0.6em]
    \footnotesize\texttt{\detokenize{#2}}}}%
  }%
}

\bibpunct{(}{)}{;}{a}{}{,}

\begin{document}

\title{Hydrodynamic and photoionization analysis of a Cosmic Dawn dwarf galaxy based on a JWST-inferred star formation history}

\author{B.~Melekh\corrauth{bohdan.melekh@lnu.edu.ua}
  \and S.~Chudyk
  \and I.~Koshmak}
\titlerunning{SED modeling of first dwarf galaxies}
\authorrunning{Melekh et al.}

\institute{Department of Astrophysics, Ivan Franko National University of Lviv,
  Kyrylo \& Methodiy str. 8, 79005 Lviv, Ukraine}
\date{}

\abstract{
We investigate the evolution of the emergent spectrum of the Cosmic Dawn dwarf galaxy 2779 using coupled hydrodynamics and photoionization modeling. Evolutionary sequences of spherically symmetric, multi-component models track the star-formation history -- derived from JWST data -- from $z \approx 29.5$ to $9.5$. BPASS stellar population synthesis is coupled with the evolution of the superwind bubble and the surrounding nebular environment. The emergent spectra include escaping stellar radiation, nebular lines and continuum, as well as X-ray emission from the hot superwind cavity. At early stages, the nebular shell suppresses the escape of Lyman-continuum (Lyc) photons, while allowing X-ray emission to emerge. Significant Lyc escape appears near $z \approx 20$, followed by a non-monotonic evolution of the escape fraction ($f_{\text{esc}}$) and the X-ray luminosity per unit star-formation rate ($L_X/\text{SFR}$). Including diffuse ionizing radiation from the nebula can result in an effective escape fraction $f_{\text{esc}} > 1$. The model sequences terminate at a stage of stagnation or the complete displacement of the outer nebular component into a shell. A comparison with JWST photometry indicates that the final model spectrum generally reproduces the shape of the observed spectral energy distribution, although the overall flux level is too high. An illustrative comparison with a previous model having a lower stellar mass indicates that accounting for the evolving contribution of the nebula can reduce the stellar mass required to match observations. However, a more precise determination of stellar mass requires a self-consistent refinement of the star formation history. These models serve as a basis for constructing representative grids of models, as well as for conducting careful checks on whether dwarf galaxies could have driven the reionization of the Universe on their own. 
}


\maketitle
\nolinenumbers

\section{Introduction} 

Recent observations of dwarf galaxies using the James Webb Space Telescope (JWST) in combination with gravitational lensing directly
indicate that these objects may provide the dominant contribution to the reionization of the Universe during the Cosmic Dawn \citep{Atek2024}.
Galaxy formation models also predict a substantial, and potentially dominant, contribution from faint dwarf galaxies to the ionizing photon budget, with the relative contribution depending on the assumed escape fraction \citep{WuKravtsov2024}.
One important limitation to a definitive confirmation of their leading role in this process is the limited size of the observational sample, since
gravitational lenses capable of probing redshifts corresponding to the Cosmic Dawn are relatively rare. Nevertheless, the results of the
aforementioned observations make the study of these objects highly relevant for modern astrophysics. One of the most topical cosmological
problems in this context is the search for spectral signatures of radiation from the first molecules in the Universe \citep{GalliPalla2013,KulinichNovosyadlyj2025}, which formed during
the Dark Ages and whose abundances undergo significant changes during the Cosmic Dawn due to the emergence of ionizing radiation \citep{Novosyadlyj2022}.

Accurately estimating the contribution of ionizing radiation from the first generations of stars requires not only a proper evolutionary
population synthesis of Population III and Population II stellar populations \citep{Schaerer2003,Zackrisson2011}, but also a reliable determination of the fraction of
ionizing photons escaping from dwarf galaxies \citep{Wise2014,Ma2015,Rosdahl2022}. Along their propagation from star-forming regions to the outer baryonic radius of a
galaxy, ionizing photons traverse the nebular medium, which both absorbs ionizing radiation and re-emits it in the form of diffuse
ionizing radiation.

A number of studies have modeled the interstellar and circumgalactic gas through which ionizing radiation propagates before escaping from early dwarf galaxies into the intergalactic medium \citep[e.g.,][]{Kimm2014,Wise2014,Ma2015,Paardekooper2015}. In this work, we focus on the SPHINX simulation, presented in \citet{Rosdahl2018}. It tracks the formation and spatial distribution of stellar particles (representing stellar populations that emit ionizing radiation), as well as the three-dimensional evolution of the gas. These cosmological radiation-hydrodynamic simulations encompass the evolution of an inhomogeneous gas distribution -- including gas accretion, galaxy mergers, stellar feedback, and the formation of low-column-density channels through which ionizing photons can escape. Stellar radiation propagates through the interstellar, circumgalactic, and intergalactic media, with the fraction of radiation escaping from each halo evaluated at its virial radius.

However, diffuse ionizing recombination radiation is treated using the on-the-spot (OTS) approximation. Instead of explicitly transporting these photons between cells, the simulation assumes that they are absorbed locally -- within the same cell where they were emitted. Specifically, this approach entails the local absorption of Lyman-continuum photons, which are produced by recombinations directly to the ground state. 

In \citet{Melekh_2024}, we demonstrated that reproducing the observed morphologies of such objects-when modeling the emission from the axisymmetric nebular environment of a dwarf galaxy with a central star-forming region -- requires a detailed treatment of diffuse ionizing radiation. Rotation shapes this environment, with "superwind" material and ionizing radiation escaping primarily along the rotation axis. However, an approximation that considers radiation propagation only in the outward direction fails to reproduce the spatial ionization structure in the equatorial regions. In the low-density conditions characteristic of the nebular environments of dwarf galaxies, the "on-the-spot" approximation --employed in the SPHINX project-proves to be even less justified.

On the other hand, \citet{Buhajenko-Melekh-2016} showed that, for spherically symmetric H{\sc ii} region models, the outward-only approximation often yields ionization structures very similar to those obtained with a detailed treatment of diffuse ionizing radiation.
Therefore, we have adopted a more gradual approach, which we consider to be more methodologically and physically consistent. The first stage involves calculating spherically symmetric multi-component photoionization models (MPhM) that account for the presence and evolution of a bubble created by a "superwind." In the subsequent stage, these calculations will be extended to evolutionary sequences of rotating dwarf galaxy models, incorporating a detailed treatment of diffuse ionizing radiation. Following this, we will analyze the differences between the results obtained during these two stages --specifically regarding the parameters $f_{\rm esc}$ and $L_X/{\rm SFR}$ -- in order to guide and optimize subsequent, more complex calculations. Here, $L_X/{\rm SFR}$ denotes the escaping X-ray luminosity
below 2~keV per unit star formation rate
(see Eq.~(8) of \citealt{Park2019}). Only then do we plan to employ a cosmological radiation-hydrodynamics strategy, developed by the authors of the SPHINX project, but incorporating a more detailed description of diffuse ionizing radiation. This paper presents the first stage of this program.

Photoionization modeling of nebular environments provides a self-consistent framework for calculating this radiative transfer while
accounting for the key physical processes in nebular plasma that both regulate and respond to ionizing radiation. In this context, the
evolutionary population synthesis models BPASS and Yggdrasil commonly recommend the use of the photoionization code Cloudy to model such
environments.

At the same time, evolutionary population synthesis models also predict the properties of stellar feedback in the form of superwinds
driven by star-forming regions, including the mechanical luminosity and mass-loss rates, both total and element-specific. These
quantities, together with the ionizing spectral energy distribution, evolve with time. This implies that adopting a static nebular medium
with a density profile determined solely by the baryonic matter distribution within a dark matter halo is inadequate, as stellar feedback
drives strong dynamical evolution of the gas. To capture this effect, it is necessary to compute a hydrodynamical model of the superwind
bubble and to perform photoionization modeling based on the resulting time-dependent structure.

Such modeling requires specifying the relationship between the halo mass and the baryonic mass and star formation
rate. In addition, determining the chemical composition of the individual components of the superwind bubble requires knowledge of the
element-specific mass-loss rates from the star-forming region. These quantities are inherently time-dependent and must therefore be
tracked consistently over the evolutionary timescale.

At present, two main approaches can be adopted to obtain the required input parameters:
\begin{enumerate}
\item by employing results from cosmological simulations of structure formation, including galaxies (e.g. Illustris);
\item by using constraints derived from modeling the star formation histories (SFH) of dwarf galaxies observed at Cosmic Dawn redshifts
($z\approx 9-11$).
\end{enumerate}

In \citet{Brummel-Smith-2023}, the star formation histories (SFHs) of dwarf galaxies observed with JWST at redshifts $z \gtrsim 10$ were modeled and
subsequently evolved down to redshifts corresponding, according to the SFH modeling, to halo mass of $\sim 10^{7}\,M_\odot$. The
authors also accounted for additional contributions to the galaxy luminosities arising from processes associated with black holes and
binary systems.

In this work, we focus exclusively on the evolution of the star-forming region and the response of the surrounding nebular environment to a superwind (a powerful galactic wind). Therefore, from a sample of 12 galaxies whose star-formation histories were modeled by \citet{Brummel-Smith-2023}, we selected object 2779: its observed photometric characteristics are well reproduced by their model without accounting for any contribution from an active galactic nucleus (AGN). This choice is consistent with our current model, which does not assume the presence of an active nucleus.

Section 2 outlines the modeling strategy. Section 3 describes the stellar population synthesis calculations used to obtain the time-dependent stellar spectra and superwind input parameters. Section 4 presents the multicomponent photoionization model (MPhM), the evolutionary sequences, the resulting thermal and ionization structures, and the comparison with JWST photometry. Finally, Section 5 summarizes our conclusions.

\section{Modeling strategy}

In this work, we investigate the evolution of the dwarf galaxy 2779 during the Cosmic Dawn using its star formation history (SFH) inferred from JWST observations. We selected this galaxy because its model-predicted stellar mass is in the best agreement with the mass inferred from James Webb Space Telescope observations. The object shows no evidence of a bright active galactic nucleus (AGN), although an AGN may still be present but obscured \citep{Brummel-Smith-2023}. We therefore focus on the evolution of the stellar ionizing radiation field without accounting for an additional ionization source from an AGN – a task that lies beyond the capabilities of current models. 

The SFH and stellar population properties of galaxy 2779 are taken from \citet{Brummel-Smith-2023}, who modeled the spectral energy distributions (SEDs) of 12 high-redshift galaxy candidates observed with JWST/NIRCam. Their framework combines halo growth, stellar population synthesis, and a possible AGN contribution. Model parameters were constrained by fitting synthetic photometry to the observed fluxes with the EMCEE ensemble Markov chain Monte Carlo (MCMC) algorithm \citep{ForemanMackey2013}.

To reconstruct the SFH, \citet{Brummel-Smith-2023} traced the host halo's evolution backward from the observed redshift to a mass of $10^7\,M_\odot$, using 5~Myr time steps and either the ellipsoidal Press--Schechter or exponential growth model. They inferred the stellar mass at each step from the stellar mass--halo mass relation of \citet{Behroozi_2013}, assuming a maximum star formation efficiency of 0.03, and calculated the star formation rate from the change in stellar mass between successive steps. Stellar metallicities were estimated using the mass--metallicity relation derived from IllustrisTNG by \citet{Torrey_2019}.

In the framework of \citet{Brummel-Smith-2023}, the resulting stellar mass and metallicity histories were used to construct the stellar SED as a sum of simple stellar populations formed at 5~Myr intervals. The models included Chabrier and standard initial mass functions (IMFs) in BPASS, and a top-heavy IMF in Yggdrasil; the BPASS models also considered both single-star and binary populations. A possible AGN contribution was represented by an additional continuum based on the SKIRTOR model \citep{Stalevski2016}. Its normalization was set by the black hole-to-stellar mass ratio, $f_{\rm BH}=M_{\rm BH}/M_\star$, while $\delta_{\rm AGN}$ controlled the slope of the optical spectrum. These parameters, together with the halo mass and dust optical depth -- and the growth-rate parameter in the exponential halo model -- were varied during the MCMC fitting.

Using the reconstructed SFH and stellar population parameters for galaxy 2779, we construct an evolutionary sequence of MPhMs for its star-forming region and surrounding nebula. At each stage, the stellar SED provides the ionizing radiation, while the time-dependent parameters describe the superwind and its interaction with the surrounding gas. The resulting sequence links the photoionization models to the galaxy's SFH rather than treating its nebular environment as a single static configuration.

Star formation in dwarf galaxies is not confined to a single central region. However, both observations of star-forming compact dwarfs and numerical simulations show that dense, low-angular-momentum gas tends to accumulate near the gravitational potential minimum, where it can form a dominant central star-forming complex. Thus, our centrally located source represents a compact, low-angular-momentum regime of star formation, rather than a necessary stage in the evolution of every dwarf galaxy. The opposite regime, in which angular momentum supports an extended disk and distributes star formation across the galactic body, would require non-spherical simulations. We plan to investigate this regime in future non-spherical models that will include gas angular momentum and a detailed treatment of diffuse ionizing radiation. 

The hydrodynamic background adopted in our MPhMs is based on the classical solutions of \citep{Weaver75} and \citep{Weaver77} for an expanding wind-driven bubble, combined with the free-wind solution of \citep{Chevalier}. It clearly separates the free-expansion superwind region, the hot thermalized cavity, the swept-up shell, and the hydrodynamically undisturbed environment. The time-dependent mechanical luminosity, mass-loss rate, superwind velocity, chemical composition, and ionizing spectral energy distribution (SED) are obtained from the corresponding population synthesis calculation. The temperatures in the first two components are given by the hydrodynamic model, while the thermal and ionization structures of the outer components are calculated using Cloudy c23.01. The outer radius, $R_4$, is constrained by the total baryonic mass in the halo and its distribution among the four components.

For modeling the radiative transfer from the region of active star formation in a dwarf galaxy through the surrounding nebular gas structures evolving under the action of a superwind, we adopt as a basis the approach previously developed by us in earlier works \citet{KM_2017,AASP_2018,Melekh25}, which has been modified here for the case of dwarf galaxies in the Cosmic Dawn epoch residing in dark matter halos of different masses. In this section, we describe the algorithm of this modified method.


\section{The evolutionary population synthesis}
\label{sec:BPASS}
To reproduce the spectral energy distribution (SED) and the mass outflows from the star-forming region in the first component of the superwind bubble, we employ the Binary Population and Spectral Synthesis (BPASS), a modern evolutionary population synthesis (EPS) model that incorporates binary stellar evolution. In this work, we use BPASS version 2.2.1. \citet{eldridge2017}

BPASS provides a grid of models covering a wide range of metallicities: $Z = 0.00001$, $0.0001$, $0.001$, $0.002$, $0.003$, $0.004$, $0.006$, $0.008$, $0.01$, $0.014$, $0.02$, $0.03$, $0.04$, as well as two types of initial mass functions (IMFs). The first IMF follows a broken power law (Standard IMF):

\begin{align}
N(M < M_{\rm max}) \propto{} & \int_{0.1}^{M_1} \left( \frac{M}{\rm M_\odot} \right)^{\alpha_1} dM \nonumber \\
& + M_1^{\alpha_1} \int_{M_1}^{M_{\rm max}} \left( \frac{M}{\rm M_\odot} \right)^{\alpha_2} dM .
\end{align}

The second IMF is the \citet{Chabrier} IMF, which BPASS implements using an exponential cutoff at the low-mass regime to reproduce the smooth transition from low-mass to high-mass stars.

\begin{table}[htbp]
\centering
\setlength{\tabcolsep}{3pt}
\caption{The range of stellar initial mass functions (IMFs) calculated in BPASS.}
\label{tab:imf}
\begin{tabular}{lcccc}
\hline\hline
Model & $\alpha_1$ & $\alpha_2$ & $M_1$ & $M_{\mathrm{max}}$ \\
 & $(0.1\,\mathrm{M}_{\odot} - M_1)$ & $(M_1 - M_{\mathrm{max}})$ & & \\
\hline
100\_100    & $-1.30$ & $-2.00$ & $0.5\,\mathrm{M}_{\odot}$ & $100\,\mathrm{M}_{\odot}$ \\
100\_300    & $-1.30$ & $-2.00$ & $0.5\,\mathrm{M}_{\odot}$ & $300\,\mathrm{M}_{\odot}$ \\
135\_100    & $-1.30$ & $-2.35$ & $0.5\,\mathrm{M}_{\odot}$ & $100\,\mathrm{M}_{\odot}$ \\
135\_300    & $-1.30$ & $-2.35$ & $0.5\,\mathrm{M}_{\odot}$ & $300\,\mathrm{M}_{\odot}$ \\
135all\_100 & $-2.35$ & $-2.35$ & $0.5\,\mathrm{M}_{\odot}$ & $100\,\mathrm{M}_{\odot}$ \\
170\_100    & $-1.30$ & $-2.70$ & $0.5\,\mathrm{M}_{\odot}$ & $100\,\mathrm{M}_{\odot}$ \\
170\_300    & $-1.30$ & $-2.70$ & $0.5\,\mathrm{M}_{\odot}$ & $300\,\mathrm{M}_{\odot}$ \\
chab100     & exp cutoff & $-2.3$ & $1.0\,\mathrm{M}_{\odot}$ & $100\,\mathrm{M}_{\odot}$ \\
chab300     & exp cutoff & $-2.3$ & $1.0\,\mathrm{M}_{\odot}$ & $300\,\mathrm{M}_{\odot}$ \\
\hline
\end{tabular}
\end{table}

BPASS offers data over an age range from $10^{6}$ to $10^{11}$ years, with a step of $0.1$ dex. For each age, the luminosity of a starburst population, $f$, is modeled over a wavelength range from $1$ to $10^{5}$ \AA.

To reconstruct the luminosity for a given star formation rate (SFR), we use the following expression:

\begin{align}
F(\lambda, t) &= \int_{0}^{t} f(\lambda, t')\,\frac{\mathrm{SFR}(t - t')}{10^{6}}\,dt' \nonumber \\
&\approx \frac{\mathrm{SFR}_{0}}{10^{6}}\, f_{0}(\lambda)\,\Delta t_{0}
+ \sum_{i=1}^{i=\max} \frac{\mathrm{SFR}_{i}}{10^{6}}\, f_{i}(\lambda)\,\Delta t_{i}
\;\; [\mathrm{erg\,s^{-1}\,\mathring{A}^{-1}}]
\label{eq:luminosity}
\end{align}
where $f$ is the luminosity of a starburst population of a given age and at a given wavelength, as provided by BPASS; SFR is the star formation rate; $t$ is the maximum age of the stellar population; and $t'$ is the instantaneous age of the stellar population. The $\Delta t_i$ values represent the time intervals (bins), defined following the BPASS recommendations. Within each time interval, the luminosity is assumed to remain constant. The first bin is defined as $\Delta t_0 = 10^{6.05}$, while all subsequent bins are defined as $\Delta t_i = 10^{6.15+0.1(i-1)} - 10^{6.05+0.1(i-1)}$.

Based on the BPASS burst models and Equation~(\ref{eq:luminosity}), we developed a dedicated code that incorporates both bursty and continuous star formation histories, as well as the evolution of metallicity, within EPS calculations. Taking the star formation histories (SFHs) from~\citet{Brummel-Smith-2023}, we reconstruct the SEDs of dwarf galaxy 2779 using the Chab100 IMF and binary stellar population models, identified as optimal for this galaxy (see \citealt{Brummel-Smith-2023}). The resulting SEDs, together with the other parameters, are then used as input for multicomponent photoionization modeling.


\section{Radiative transfer through the nebular components}
\label{sec:MPhM}


\subsection{Task overview}

Before entering the intergalactic medium, ionizing photons originating from the stellar photospheres within the active star-forming
region of a dwarf galaxy propagate through the surrounding nebular environment. This environment may exhibit a rather complex density
distribution, primarily due to the expansion of a superwind bubble formed and sustained by the star-forming region. Even under the
assumption of spherical symmetry adopted here, owing to the lack of sufficient information on the angular momentum of the first dwarf
galaxies, the bubble expands into an inhomogeneous medium: the gas density within the halo remains approximately constant only within a
relatively small central region and decreases with increasing distance from the center.

The chemical composition of the nebular environment is also expected to be inhomogeneous. Outside the superwind bubble, the chemical
composition corresponds to the mean abundance pattern appropriate for a given redshift. In the inner part of the bubble, however, it is
determined mainly by the metallicity of stars that are currently evolving or have already completed their evolution. In the outer part of
the bubble, near the contact discontinuity, the chemical composition results from the mixing of these two components.

Moreover, the nebular medium itself constitutes a source of ionizing photons. Diffuse ionizing radiation is generated by processes in the
nebular plasma induced by the stellar ionizing radiation field. Additional emission arises from gas heated by the propagation of
superwind shocks, particularly the reverse shock. All these processes modify the spectral energy distribution (SED) of the stellar
ionizing radiation as it propagates through the nebular environment. Consequently, the ionizing SED emerging from the dwarf galaxy into
the intergalactic medium may differ substantially from the intrinsic SED produced by the superposition of the radiation from all stars in
the star-forming region at a given evolutionary time. Therefore, this transformation must be taken into account when determining the
emergent radiation spectrum of a dwarf galaxy.

In this work, we adopt the approach developed in our previous studies, an overview of which can be found in \citet{Melekh25}. Our approach
is based on the models of \citet*{Weaver75} and \citet{Chevalier}, which can be regarded as classical studies of the physical
conditions within stellar winds and superwinds, respectively. The results of these models can be readily expressed in a self-similar
form, which facilitates the implementation of the corresponding hydrodynamical solutions into a photoionization code.

Clearly, by adopting these models, we restrict our calculations to the corresponding evolutionary phase of superbubble expansion, without
explicitly verifying whether this approximation remains applicable at each evolutionary time considered. Developing criteria for tracking
the applicability of the adopted evolutionary regime is a separate problem that we are also currently investigating. However, as will be
shown below, the stopping criteria adopted in our calculations prevent the models from extending over excessively long evolutionary
intervals. This allows us to assume that our calculations remain within the longest-lasting evolutionary phase of the superbubble
considered by the adopted model.

In addition, our calculations account for the time dependence of both the superwind input parameters and the density of the ambient
medium into which the bubble expands. This required modifications to the aforementioned models. For these reasons, we present our
modeling approach in considerable detail below.

\subsection{Model components and modeling algorithm}

\begin{figure}
    \centering
    \includegraphics[width=0.48\textwidth]{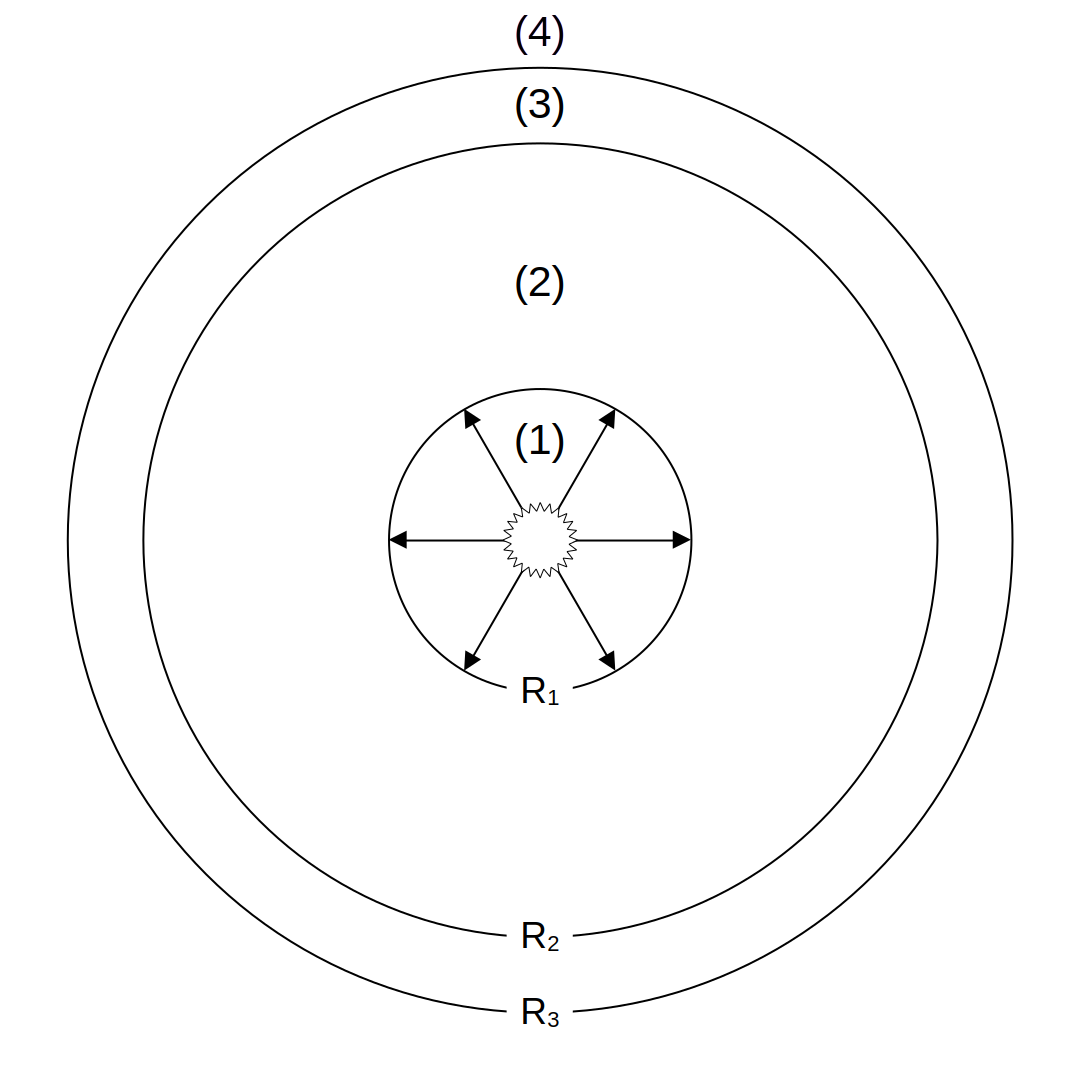}
    \caption{Superwind bubble and its components (see text).}
    \label{fig:Bubble}
\end{figure}

We divide the nebular environment surrounding the region of active star formation into the following four components, ordered radially from the
center outward (see Fig.~\ref{fig:Bubble}):
\begin{enumerate}
\item the free-expansion region of the superwind;
\item the superwind cavity, filled with rarefied gas thermalized by the reverse shock of the superwind at radius $R_1$, which separates
the first and second components;
\item a relatively thin, high-density shell swept up by the forward shock; the contact discontinuity separating the second and third
components is located at radius $R_2$;
\item the hydrodynamically unperturbed ambient environment, which is predominantly photoionized (an H{\sc~ii} region) by both stellar
ionizing radiation and diffuse ionizing photons.
\end{enumerate}

In our models, the temperature distribution in the first component corresponds to the solutions by Chevalier–Clegg model \citet{Chevalier}.
Taking into account the data from \citet{Weaver77}, the velocity in the first component corresponds to the stellar superwind velocity.
The density in the first component was calculated following \citet{Weaver75}:
\begin{equation}
n_1 = \frac{\dot{M}_w}{4 \pi r^2 \, ChF_{in} \, m_{\mathrm{H}} \, v_w} \ \mathrm{[cm^{-3}]},
\end{equation}
where $r$ is the distance from the center of the star-forming region, $m_{\text{H}}$ is the mass of a hydrogen atom, and $ChF_{in}$ is the
total abundance of all considered chemical elements relative to hydrogen for the component (1).

The chemical composition in the first component was determined using the results of modeling obtained with BPASS \citet{BPASS}.
From BPASS we obtained the mass-loss rate of the star-forming region due to the stellar wind and supernova explosions, $\dot{M}$, 
as well as the individual mass-loss rates of hydrogen, $\dot{H}$, and helium, $\dot{He}$, expressed in solar masses per year.
The mass fractions of hydrogen, helium, and heavy elements were determined as follows:
\begin{equation}
X_{\mathrm{H}} = \dot{M}_{\mathrm{H}} / \dot{M}; \quad
X_{\mathrm{He}} = \dot{M}_{\mathrm{He}} / \dot{M}; \quad
Z = 1 - X_{\mathrm{H}} - X_{\mathrm{He}}.
\end{equation}
According to \citet{Byrne2025}, the mass fraction as well as relative abundances of heavy elements were calculated for 
$[\alpha/{\rm Fe}]=+0.2$. Then, total abundance was calculated as follows:
\begin{equation}
\begin{aligned} 
ChF_{\mathrm{in}} =\; & 1 
+ 4\,\frac{\mathrm{He}}{\mathrm{H}} 
+ 12\,\frac{\mathrm{C}}{\mathrm{H}} 
+ 14\,\frac{\mathrm{N}}{\mathrm{H}} 
+ 16\,\frac{\mathrm{O}}{\mathrm{H}} 
+ 20\,\frac{\mathrm{Ne}}{\mathrm{H}} \\
& + 24\,\frac{\mathrm{Mg}}{\mathrm{H}} 
+ 28\,\frac{\mathrm{Si}}{\mathrm{H}} 
+ 32\,\frac{\mathrm{S}}{\mathrm{H}} 
+ 36\,\frac{\mathrm{Cl}}{\mathrm{H}} \\
& + 40\,\frac{\mathrm{Ar}}{\mathrm{H}} 
+ 56\,\frac{\mathrm{Fe}}{\mathrm{H}}.
\label{formula:ChFin}
\end{aligned}
\end{equation}

The radius describing the outer boundary of component (1) and the inner boundary of component (2) actually represents the reverse shock location and
is defined as follows:
\begin{equation}
R_1(t) = 7.02 \left[ \frac{4}{3} \int_0^t \frac{L(t)\, v(t)^{-5/3}\, t^{1/3}}{n\bigl(R_3(t)\bigr)} \, dt \right]^{0.3} \ \mathrm{[pc]},
\label{formula:R1}
\end{equation}
where $L(t)$ -- mechanical luminosity, $v(t)$ -- superwind velocity, and $n(R_3)$ -- gas density at the outer side of $R_3(t)$.

The contact discontinuity radius, $R_2(t)$ as a function of time was calculated in the following way:
\begin{equation}
R_2(t) = 31 \left[ 3 \int_0^t \frac{L(t)\, t^2}{n\bigl(R_3(t)\bigr)} \, dt \right]^{0.2} \ \mathrm{[pc]}.
\label{formula:R2}
\end{equation}

The procedure to obtain the temperature, density and velocity distributions in component 2 from the hydrodynamical model by \citet{Weaver77}
taking into account gas evaporation from component 3 into component 2 due to thermal conductivity is described in detail in our work
\citet{Melekh25}. Here we note that formulas in this paper were modified to adapt
them to time-dependent parameters of superwind and a non-uniform radial distribution of baryonic matter in the halo. 
The chemical composition of component 2 is adopted as a mixture of chemical compositions of component 1 and component 3.
The procedure of calculation of these abundances is also described in \citet{Melekh25}.

The density was not assumed to be constant throughout component 3. Instead, we adopted a distribution that accounts for evaporation from
the inner side of component 3, and it increases with radius:
\begin{equation}
n_3(r) = n(R_2) + [n(R_3)-n(R_2)]\left(\frac{r-R_2}{R_3-R_2}\right)^{0.3},
\label{formula:n3}
\end{equation}
where $R_3$ corresponds to the outer radius of the 3rd component. 
The value of $n(R_2)$, density at $R_2$ at the side of component 3 is evaluated iteratively during modeling to match the pressure on the component-2 side of $R_2$, because of contact discontinuity at this radius.
Because we do not know a priori the value of $R_3$ required to calculate distribution of $n_3(r)$, it was necessary
to perform the global iterations for each model on the basis of the mass of component 3, which is known a priori, until the values of $R_3$ obtained in successive iterations agree within the adopted numerical tolerance.  This mass corresponds to mass of all
baryonic matter that was at bubble location before it expanded but corrected for the mass of gas evaporated into component 2. Such prior 
distribution followed the same radial distribution as in component 4 that corresponds to the baryonic matter distribution in halo
 (see \citet{Nebrin2023}):
\begin{equation}
\label{formula:n4}
n_4(r) = \frac{n_{core}}{1 + \left(\frac{r}{R_{core}}\right)^2},
\end{equation}
where $n_{core}$ -- gas density within $R_{core}\approx 0.1 R_{vir}$, $R_{vir}$ -- virial radius of the galaxy calculated by us as in
\citet{Nebrin2023}.

\begin{figure*}
\centering
\includegraphics[width=1.00\textwidth]{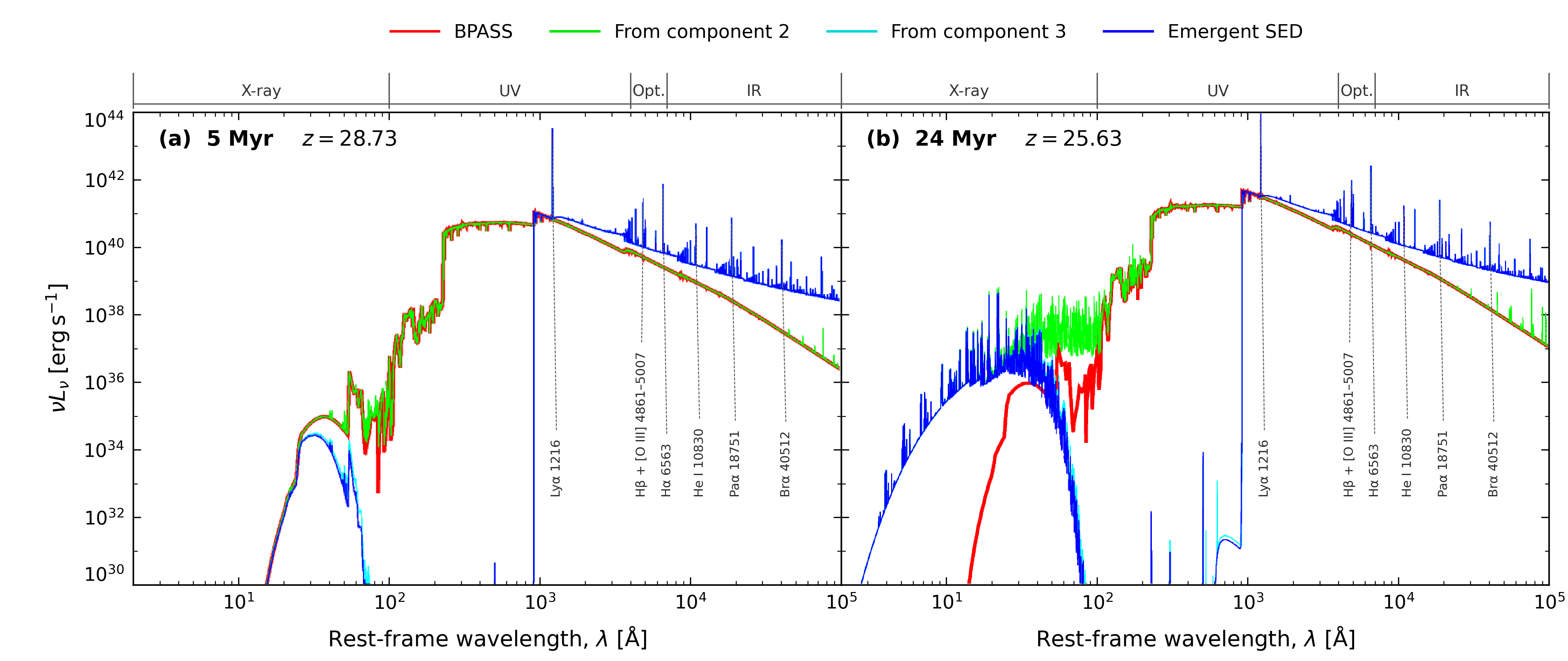}
\caption{The evolution at early stages of SFH of the stellar spectrum -- calculated using the BPASS evolutionary stellar population synthesis method -- as radiation propagates through the components of a superwind bubble and the hydrodynamically undisturbed ambient medium of the nebular environment. The resulting spectrum is calculated at the outer baryonic boundary of the galaxy ($R_4$).}
\label{SEDs524}\label{fig:SEDs_5_24}
\end{figure*}

\subsection{Evolutionary sequences of MPhMs}

One of the most pressing questions in the study of dwarf galaxies during the Cosmic Dawn is whether they are the primary sources responsible for cosmic reionization. Consequently, a key result of MPhM is the fraction of ionizing photons that escape a dwarf galaxy at various stages of its evolution within a dark matter halo. Also of interest is the evolution of the ionizing spectrum as stellar radiation propagates through the surrounding nebular medium. In our models, this medium encompasses the central region of active star formation within a halo of a given mass; it consists of components 1-3, associated with a superwind bubble, as well as an outer, hydrodynamically undisturbed nebular component 4. We define the ionizing photon escape fraction ($f_{\rm esc}$) as the ratio of the total flux of ionizing photons crossing the outer boundary of component 4 ($R_4$) to the total number of ionizing photons emitted by all stars within the galaxy.

Our MPhM evolutionary sequences are computed with an age step of 1~Myr consistent with the temporal sampling of the EPS input data, which are based on the SFH from \citet{Brummel-Smith-2023}. Each individual MPhM calculation is terminated upon reaching the outer boundary of the galaxy's baryonic component ($R_4$). This radius is determined in a self-consistent manner: the total mass of stars and gas contained within the model components is required to equal a specified total baryonic mass of the galaxy. This mass balance is monitored throughout the entire calculation process.

Each stage of an evolutionary model sequence is terminated when one of the following criteria is met:
\begin{enumerate}
\item A stagnation radius is reached. In our models, this occurs when the pressure at the inner boundary of component 4 becomes equal to or exceeds the pressure at the outer boundary of component 3. We assume that, following this, the bubble begins to contract rapidly; therefore, the calculation of the current expansion phase is terminated. The subsequent model in the temporal sequence is treated as the beginning of a new expansion phase.
\item Component 4 is completely swept into Component 3. At this stage, the outer baryonic boundary of the galaxy is defined by the radius $R_3$ rather than $R_4$. The distribution of gas in the region between this radius and the halo's virial radius ($R_{\rm vir}$) -- which is typically 2--2.5 times larger than the outer baryonic radius -- remains uncertain. While this region is expected to contain intergalactic gas, uncertainties in modeling its accretion onto the halo would introduce significant additional error into our models. Therefore, we cease our analysis of the model sequence at the point where Component 4 disappears. If we assume a sharp drop in gas density at the outer baryonic boundary -- provided that the superwind intensity does not diminish -- the shell corresponding to Component 3 will expand much more rapidly, leading to a reduction in its optical depth to ionizing radiation. Given these assumptions, the $f_{\rm esc}$ values derived from such constrained model sequences should be regarded as lower limits if they are used as input astrophysical parameters in cosmological calculations.
\end{enumerate}

Models within the redshift range from $z\approx 29.5$ (corresponding to zero age along the adopted SFH of the galaxy) to $z\approx 20$ correspond to low-mass halos ($M_h = (1-40) \cdot 10^7\ M_\odot$) containing approximately $(1-60)\cdot 10^6\  M_\odot$ of gas. In such systems, a superwind bubble expands and, within 4 -- 5 Myr, completely displaces component 4 into component 3. This triggers the second criterion for terminating the simulation. The resulting dense shell (component 3) blocks nearly all ionizing radiation originating from the central star-forming region. Figure \ref{fig:SEDs_5_24} presents the spectra originating from the baryonic components of these galaxies at early stages of their evolution -- specifically at 5 Myr (z=28.7) and 24~Myr (z=25.6). The model corresponding to an age of 24~Myr represents the final calculated age within one of the early expansion cycles of the bubble, just before it ceases expansion upon reaching the stagnation radius. The spectra span the X-ray, UV, optical, and IR ranges, extending to a wavelength of $10^5$\AA, which corresponds to the low-energy limit of the BPASS spectra. Results are presented for selected ages (see text) based on the star formation history of galaxy 2779 from \citet{Brummel-Smith-2023}. The corresponding redshifts and the outer radii of the model components for which the spectra were calculated are also given. A comparison of the radius $R_4$ with the virial one ($R_{\rm vir}$) suggests that the baryonic matter is deeply embedded within a dark matter halo. Additionally, the fraction of ionizing photons escaping beyond the baryonic component of the galaxy, calculated at the radius $R_4$, is provided.

\begin{figure*}
\centering
\includegraphics[width=1.00\textwidth]{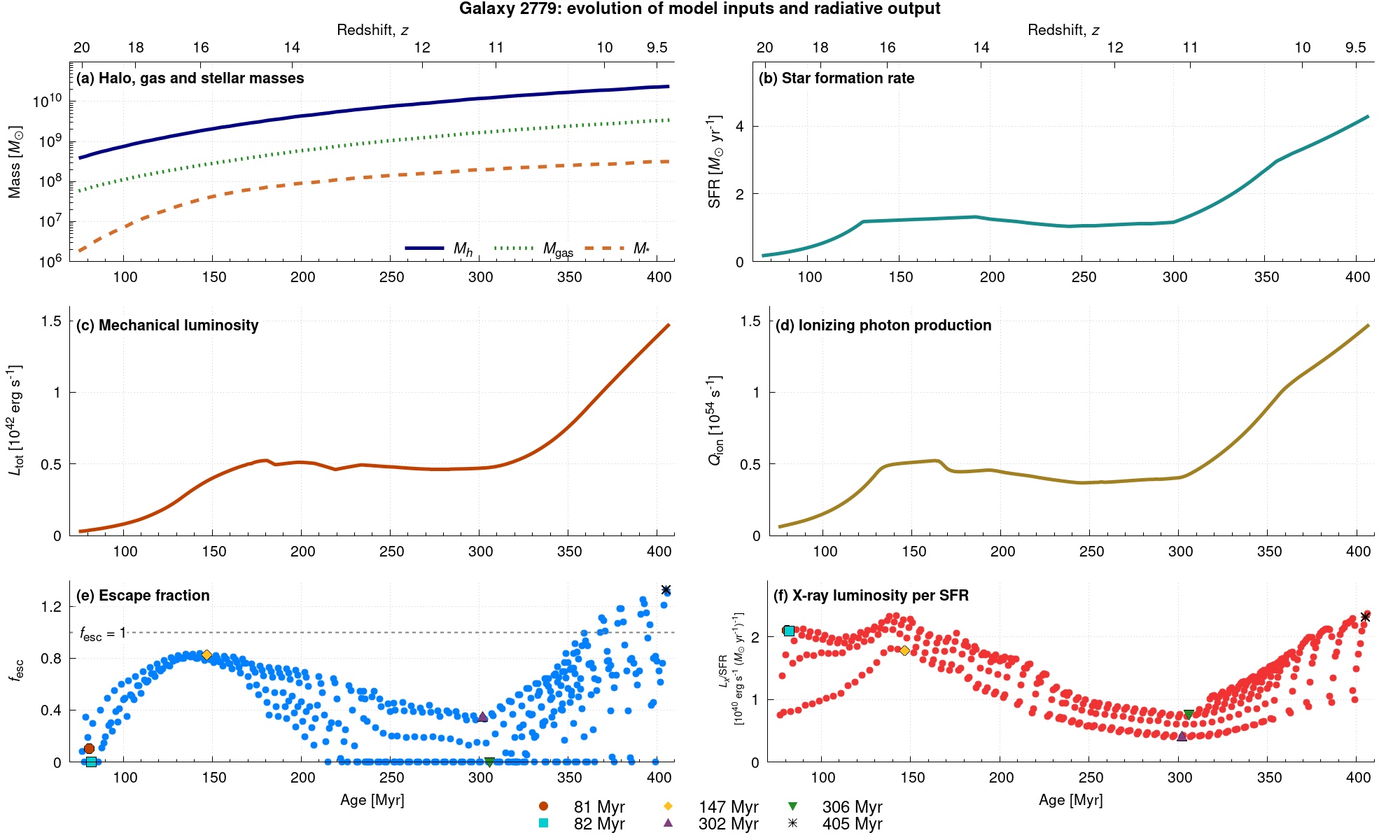}
\caption{Evolution of the MPhM 
input parameters, including the halo mass, $M_h$; the baryonic masses in gas, $M_g$, and stars, $M_*$, shown separately; the star formation rate, SFR; the total mechanical luminosity, $L_{\rm tot}$; and the ionizing photon production rate, $Q_{\rm ion}$, together with the model outputs: the escape fraction of ionizing radiation, $f_{\rm esc}$, and the X-ray luminosity normalized by the star formation rate, $L_X/{\rm SFR}$.}
\label{fig:Inp_Out}
\end{figure*}

At redshifts of 28.73 and 25.63, a noticeable drop in spectral luminosity at wavelengths $\lambda\leq 912\text{\AA}$ indicates that the surrounding nebular medium is almost completely opaque to ionizing radiation in UV. Although the emergent Lyc-radiation is almost entirely suppressed, X-rays propagate through the galaxy’s nebular medium much more effectively. At an age of 5~Myr, the high-energy wing of the X-ray feature in the emergent spectrum (on the left side of the figures) mirrors the shape of the BPASS stellar spectrum; it should be noted that close binary systems make a significant contribution to the emission in this energy range. However, at 24 Myr, a substantial additional contribution arises from a superwind cavity -- filled with hot, diffuse, thermalized gas -- which has grown significantly larger than it was at 5 Myr. Also, despite the presence of an X-ray "hump" at such high redshifts, its integrated luminosity is four to five orders of magnitude lower than the luminosity of the stellar spectrum predicted by the BPASS model. Consequently, $f_{\rm esc}$ remains close to zero.
In contrast, the non-ionizing portion of the emergent spectrum at these ages and redshifts remains relatively bright compared to the input spectrum derived via stellar population synthesis. This radiation does not ionize hydrogen or helium in their ground states, but it can affect molecular spectra in the intergalactic medium. 

Significant escape of ionizing radiation from the system begins at $z\approx 20$. Therefore, we focus our analysis mainly on simulation results for the range from $z\approx 20$ to $z\approx 9.5$ -- with the latter value corresponding to the observed redshift of galaxy 2779. 
Figure \ref{fig:Inp_Out} shows the evolution of the input parameters of the MPhM model for galaxy 2779: the halo mass, $M_h$; the baryonic masses in gas, $M_g$, and stars, $M_*$, shown separately; the star formation rate, SFR; the total mechanical luminosity, $L_{\rm tot}$; and the ionizing photon production rate, $Q_{\rm ion}$, together with the model outputs: the escape fraction of ionizing radiation, $f_{\rm esc}$, and the X-ray luminosity normalized by the star formation rate, $L_X/{\rm SFR}$. While the MPhM input parameters (panels a--d) vary relatively smoothly, the radiative output exhibits distinct branches and abrupt transitions. The ages highlighted by different symbols mark the onset of the rise in the output quantities (panels e and f), regions near local maxima and minima of $f_{\rm esc}$, and the final model, which corresponds to the redshift at which dwarf galaxy 2779 is observed. The rise in the input parameters during the final phase of the SFH also contributes to the increase in the output quantities at this stage. For these selected models, we analyze how their spectra are transformed as radiation propagates from the stars to the outer baryonic radius of the galaxy, $R_4$.

\subsection{Ionization structure of the nebular environment}

We investigate the radial distributions of electron temperature $T_e$, electron density $n_e$, and ionic fractions in the MPhMs of galaxy 2779 at ages of 5, 24, 81, 82, 147, 302, 306, and 405~Myr. These models encompass the onset of significant Lyc-radiation escape, demonstrate states near local extrema of the quantities $f_{\rm esc}$ and $L_X/{\rm SFR}$, and represent the final stage corresponding to the observed galaxy. The objective of this work is to establish a relationship between variations in the emergent spectra and the physical conditions within the four components of the nebular medium.

Temperature distributions in the freely expanding stellar wind (Component 1) and the hot, thermalized cavity (Component 2) are determined based on adopted hydrodynamic solutions. The thermal and ionization structure of the swept-up shell (Component 3) and the surrounding, hydrodynamically undisturbed gas (Component 4) are calculated using the \textsc{Cloudy} code. This distinction is important for interpreting temperature contrasts between the components.

\begin{figure*}[t]
\centering
\includegraphics[width=\textwidth]{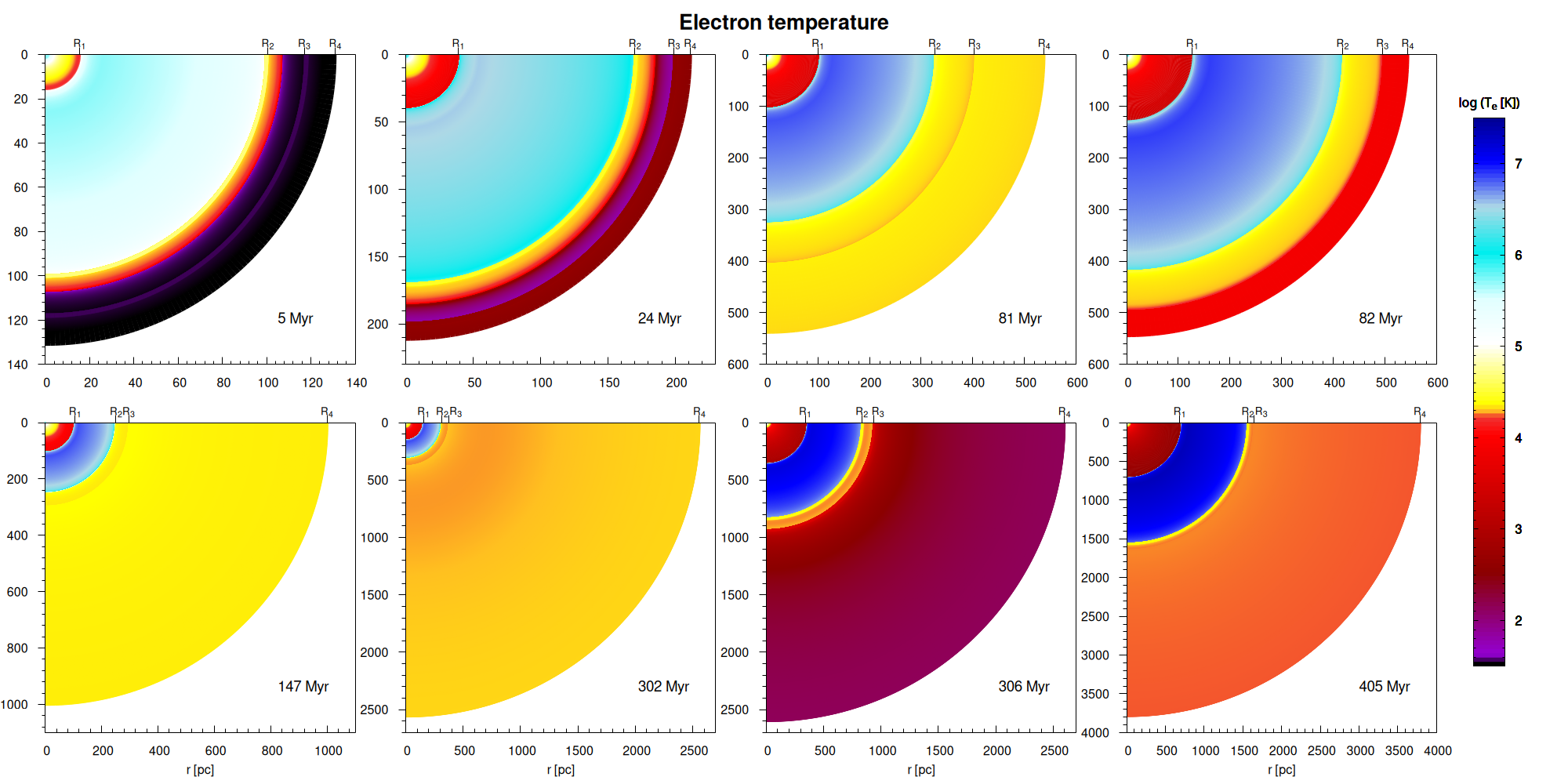}
\caption{
Electron temperature distributions in the modeled nebular environment of Galaxy 2779 at ages 5, 24, 81, 82, 147, 302, 306, and 405~Myr. The concentric images represent 
spherically symmetric radial solutions. The outer radii of the four components are denoted as $R_1$, $R_2$, $R_3$, and $R_4$; temperatures are displayed using a logarithmic color scale.}
\label{Te_bubble}
\end{figure*}

\begin{figure*}[t]
\centering
\includegraphics[width=\textwidth]{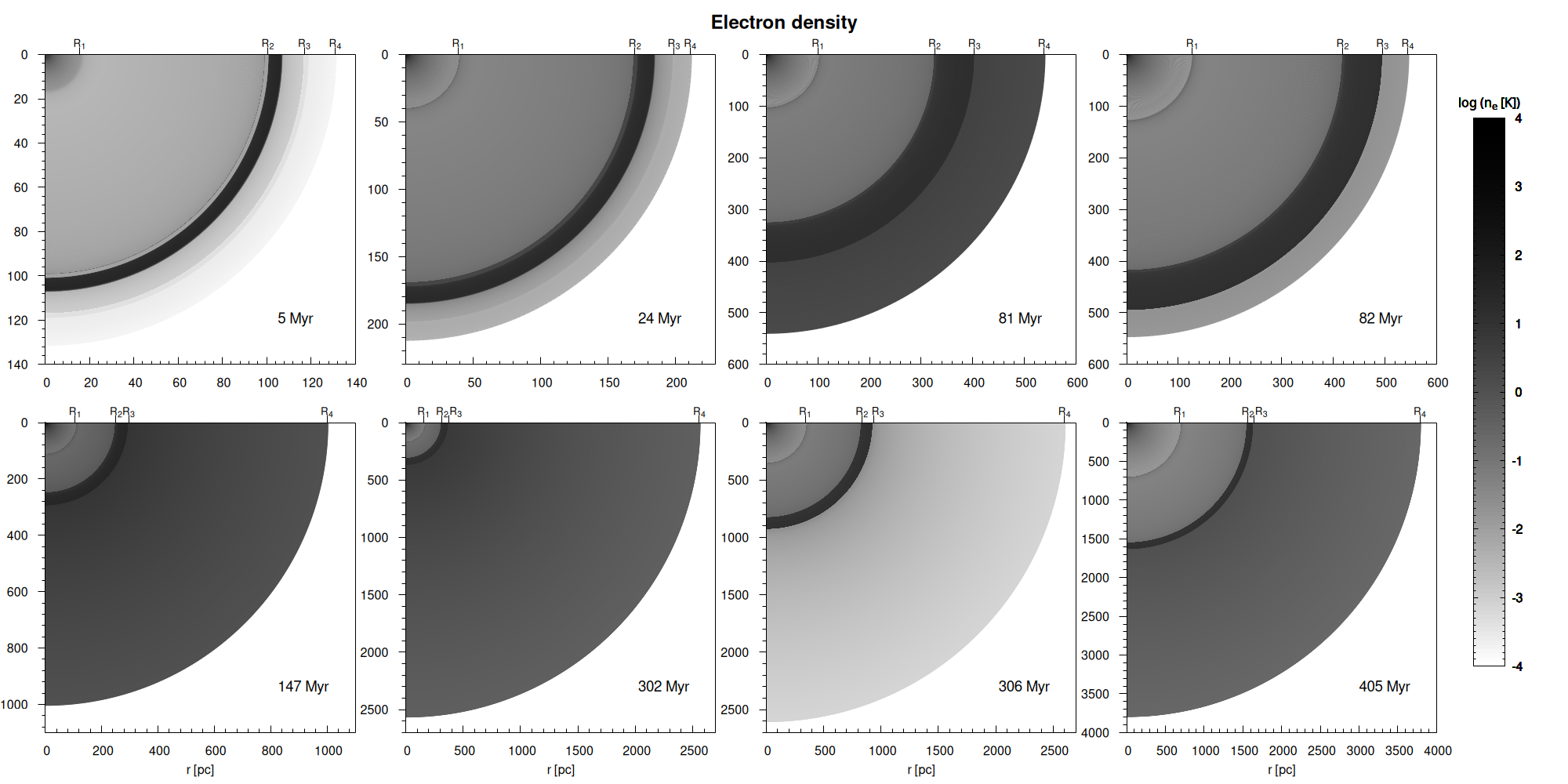}
\caption{
Electron density distributions in the modeled nebular environment of galaxy 2779 for the same ages as in Fig.~\ref{Te_bubble}. The concentric images show spherically symmetric radial solutions, with the boundaries of the components marked as $R_1$, $R_2$, $R_3$, and $R_4$. Electron densities are displayed using a logarithmic grayscale.}
\label{ne_bubble}
\end{figure*}

Figures~\ref{Te_bubble} and~\ref{ne_bubble} show the distributions of temperature and electron density as concentric plots for spherically symmetric models; the boundaries of the components are denoted as $R_1$, $R_2$, $R_3$, and $R_4$. These plots visualize radial solutions and do not imply two-dimensional calculations. The ranges of values (represented by color or shades of gray) are log~($T_e$~[K])=1.5 -- 7.5 and log~($n_e$~[cm$^{-3}$]) = -4 -- 4, respectively. Table~\ref{tab_IonStr} lists the extreme values for each component, as well as its outer radius.

Profiles of $T_e$, $n_e$, as well as the ionic fractions of hydrogen, helium, and oxygen for each component, are available at the following link:
\url{https://physics.lnu.edu.ua/wp-content/uploads/Images\_for\_archive.zip}
In these detailed profiles, the radial coordinate is measured from the inner boundary of the respective component, while $R_1$--$R_4$ denote distances from the center of the galaxy. The provided oxygen fractions cover ions ranging from O$^{0}$ to O$^{8+}$.

Two internal components remain highly ionized throughout all the evolutionary stages considered: hydrogen is almost completely ionized, while helium is predominantly doubly ionized. However, the temperature and density distributions within them evolve differently. In Component 1, the maximum temperature reaches 3.31$\cdot$10$^6$~K at 147~Myr, subsequently decreasing to 2.13$\cdot$10$^5$~K at 405~Myr. The maximum electron density increases from 3.99 to 7.72$\cdot$10$^3$~cm$^{-3}$ between 5 and 405~Myr, whereas the minimum density 
exhibits non-monotonic variation. Consequently, the evolution cannot be characterized as a uniform increase or decrease of any of these quantities within this component.

\begin{table*}[t]
\caption{Minimum and maximum electron temperatures and their densities, as well as the outer radii of the four model components of galaxy 2779 at selected ages.}
\centering
\begingroup
\scriptsize
\setlength{\tabcolsep}{3pt}
\begin{tabular}{l*{8}{c}}
\hline
Parameter &5 [Myr] &24 [Myr] &81 [Myr] &82 [Myr] &147 [Myr] &302 [Myr] &306 [Myr] &405 [Myr]\\
\hline
\multicolumn{9}{c}{Component 1}\\
\hline
$T_e^{\mathrm{min}}$ [K] &1.38$\cdot 10^{4}$ &5.76$\cdot 10^{3}$ &3.16$\cdot 10^{3}$ &2.27$\cdot 10^{3}$ &6.65$\cdot 10^{3}$ &4.17$\cdot 10^{3}$ &7.82$\cdot 10^{2}$ &3.92$\cdot 10^{2}$ \\
\hline
$T_e^{\mathrm{max}}$ [K] &5.11$\cdot 10^{5}$ &7.48$\cdot 10^{5}$ &1.06$\cdot 10^{6}$ &7.49$\cdot 10^{5}$ &3.31$\cdot 10^{6}$ &2.63$\cdot 10^{6}$ &3.85$\cdot 10^{5}$ &2.13$\cdot 10^{5}$ \\
\hline
$n_e^{\mathrm{min}}$ [cm$^{-3}$] &1.75$\cdot 10^{-2}$ &1.88$\cdot 10^{-2}$ &2.55$\cdot 10^{-2}$ &2.27$\cdot 10^{-2}$ &9.86$\cdot 10^{-2}$ &5.99$\cdot 10^{-2}$ &2.33$\cdot 10^{-2}$ &1.32$\cdot 10^{-2}$ \\
\hline
$n_e^{\mathrm{max}}$ [cm$^{-3}$] &3.99 &38.73 &3.45$\cdot 10^{2}$ &3.66$\cdot 10^{2}$ &1.44$\cdot 10^{3}$ &1.76$\cdot 10^{3}$ &3.30$\cdot 10^{3}$ &7.72$\cdot 10^{3}$ \\
\hline
$R_1$ [pc] &15.73 &38.51 &99.94 &126.73 &105.70 &155.78 &339.69 &685.20 \\
\hline
\multicolumn{9}{c}{Component 2}\\
\hline
$T_e^{\mathrm{min}}$ [K] &1.51$\cdot 10^{4}$ &7.43$\cdot 10^{4}$ &1.43$\cdot 10^{5}$ &2.21$\cdot 10^{5}$ &1.44$\cdot 10^{5}$ &1.48$\cdot 10^{5}$ &3.01$\cdot 10^{5}$ &4.33$\cdot 10^{5}$ \\
\hline
$T_e^{\mathrm{max}}$ [K] &4.91$\cdot 10^{5}$ &3.37$\cdot 10^{6}$ &6.48$\cdot 10^{6}$ &7.21$\cdot 10^{6}$ &6.51$\cdot 10^{6}$ &6.69$\cdot 10^{6}$ &1.37$\cdot 10^{7}$ &1.96$\cdot 10^{7}$ \\
\hline
$n_e^{\mathrm{min}}$ [cm$^{-3}$] &2.96$\cdot 10^{-3}$ &3.84$\cdot 10^{-2}$ &6.11$\cdot 10^{-2}$ &4.21$\cdot 10^{-2}$ &0.22 &0.16 &6.66$\cdot 10^{-2}$ &4.44$\cdot 10^{-2}$ \\
\hline
$n_e^{\mathrm{max}}$ [cm$^{-3}$] &2.58$\cdot 10^{-2}$ &1.75 &2.77 &0.35 &10.04 &7.43 &3.02 &2.01 \\
\hline
$R_2$ [pc] &98.98 &169.49 &326.38 &418.41 &250.79 &316.90 &833.82 &1561.91 \\
\hline
\multicolumn{9}{c}{Component 3}\\
\hline
$T_e^{\mathrm{min}}$ [K] &32.06 &53.70 &1.69$\cdot 10^{4}$ &1.39$\cdot 10^{2}$ &2.10$\cdot 10^{4}$ &1.77$\cdot 10^{4}$ &86.72 &1.71$\cdot 10^{4}$ \\
\hline
$T_e^{\mathrm{max}}$ [K] &4.63$\cdot 10^{4}$ &2.77$\cdot 10^{4}$ &3.01$\cdot 10^{4}$ &2.87$\cdot 10^{4}$ &3.10$\cdot 10^{4}$ &2.54$\cdot 10^{4}$ &2.42$\cdot 10^{4}$ &2.22$\cdot 10^{4}$ \\
\hline
$n_e^{\mathrm{min}}$ [cm$^{-3}$] &9.42$\cdot 10^{-4}$ &7.14$\cdot 10^{-3}$ &3.74 &1.31$\cdot 10^{-2}$ &12.21 &9.96 &1.45$\cdot 10^{-2}$ &9.19 \\
\hline
$n_e^{\mathrm{max}}$ [cm$^{-3}$] &26.84 &19.65 &13.82 &11.93 &28.46 &14.41 &10.62 &10.91 \\
\hline
$R_3$ [pc] &115.94 &198.04 &403.06 &495.07 &296.03 &377.97 &935.51 &1646.93 \\
\hline
\multicolumn{9}{c}{Component 4}\\
\hline
$T_e^{\mathrm{min}}$ [K] &30.08 &3.18$\cdot 10^{2}$ &2.00$\cdot 10^{4}$ &6.90$\cdot 10^{3}$ &2.13$\cdot 10^{4}$ &1.78$\cdot 10^{4}$ &1.31$\cdot 10^{2}$ &1.65$\cdot 10^{4}$ \\
\hline
$T_e^{\mathrm{max}}$ [K] &34.80 &4.23$\cdot 10^{2}$ &2.14$\cdot 10^{4}$ &7.10$\cdot 10^{3}$ &2.25$\cdot 10^{4}$ &1.98$\cdot 10^{4}$ &3.38$\cdot 10^{3}$ &1.74$\cdot 10^{4}$ \\
\hline
$n_e^{\mathrm{min}}$ [cm$^{-3}$] &1.62$\cdot 10^{-4}$ &3.93$\cdot 10^{-3}$ &1.60 &1.44$\cdot 10^{-2}$ &0.99 &0.37 &6.79$\cdot 10^{-4}$ &0.23 \\
\hline
$n_e^{\mathrm{max}}$ [cm$^{-3}$] &4.00$\cdot 10^{-4}$ &5.22$\cdot 10^{-3}$ &2.74 &2.04$\cdot 10^{-2}$ &7.21 &4.96 &2.10$\cdot 10^{-2}$ &0.98 \\
\hline
$R_4$ [pc] &129.13 &209.34 &537.26 &543.77 &999.90 &2558.16 &2598.59 &3784.86 \\
\hline
\end{tabular}
\endgroup
\label{tab_IonStr}
\end{table*}

Component 2 contains a hotter cavity in the final model than in the initial one: its maximum temperature rises from 4.91$\cdot$10$^5$~K (at $t$ = 5~Myr) 
to 1.96$\cdot$10$^7$~K (at $t$ = 405~Myr). The electron density within it varies non-monotonically, with the highest minimum and maximum values observed at $t$ = 147~Myr. The temperature and radial extent of this hot gas support its contribution to the emergent X-ray emission. However, the evolution of this contribution depends on a combination of factors -- temperature, density, the volume of the emitting region, and the absorption of radiation by the surrounding gas -- so an increase in cavity size alone does not uniquely 
determine the change in $L_X$ or $L_X/\mathrm{SFR}$.

Oxygen profiles allow for further differentiation of the internal components. In Component 1, the dominant ionization state (among those presented) is O$^{4+}$ (at 5, 24, and 306~Myr), O$^{5+}$ (at 81, 82, 147, and 302~Myr), and O$^{3+}$ (at 405~Myr). 

Inside the hot cavity (component 2), highly ionized oxygen predominates. The dominant ionization stage varies with radius, ranging from O$^{6+}$ to O$^{3+}$ at 5~Myr and from O$^{8+}$ to O$^{6+}$ at 24~Myr, whereas O$^{8+}$ dominates at the other selected ages.

The most pronounced changes regarding the escape of ionizing radiation occur in Components 3 and 4. At 5 and 24~Myr, Component 3 features an ionized inner region dominated by H$^{+}$, He$^{+}$, and O$^{2+}$ ions, followed by a transition to predominantly neutral gas. An enhanced abundance of O$^{+}$ ions is observed 
in the transition zone. This transition is accompanied by a sharp drop in both electron temperature ($T_e$) and electron density ($n_e$). At this age, Component 4 is also predominantly neutral. The presence of such outer neutral layers explains the strong suppression of the escaping ionizing radiation -- despite the high degree of gas ionization in the inner components and the presence of X-ray emission.

At 81, 147, 302, and 405~Myr, ions H$^{+}$, He$^{+}$, and O$^{2+}$ dominate in Component 4, while Component 3 remains predominantly ionized. In these models, 
the shell temperature remains approximately 10$^4$ K throughout; the extended, cold outer regions observed in models with suppressed Lyc-emission are absent here. The significant fraction of O$^{2+}$ ions in these photoionized components points to regions that contribute to the nebular oxygen emission considered in the spectral analysis. However, ionic fractions alone do not determine the luminosity of the corresponding spectral lines.

A comparison of the states at 81 and 82~Myr demonstrates how significantly the external ionization structure can change between closely spaced evolutionary stages. At 82~Myr, a transition to neutral gas is observed near the outer boundary of Component 3, while Component 4 becomes predominantly neutral. The minimum temperature of the shell drops from 1.69$\cdot$10$^4$~K to 1.39$\cdot$10$^2$~K, and the minimum electron density decreases from 3.74 to 1.31$\cdot$10$^{-2}$~cm$^{-3}$. A hot cavity persists in both models. These differences are consistent with a significant reduction in the ionizing radiation escape fraction ($f_{\rm esc}$), which occurs without a corresponding decrease in X-ray output.

A similar contrast is observed in the interval from 302 to 306~Myr. The minimum temperature of the shell drops from 1.77$\cdot$10$^4$~K to 86.72~K, while the minimum electron density decreases from 9.96 to 1.45$\cdot$10$^{-2}$~cm$^{-3}$. Component 4 becomes predominantly neutral once again, whereas the maximum temperature of the cavity rises from 6.69$\cdot$10$^6$~K to 1.37$\cdot$10$^7$~K. The coexistence of a hot cavity and neutral outer layers explains why, in the model at 306~Myr, the leakage of ionizing radiation can be negligible, while X-ray emission remains significant. The extremely low temperature values apply only to the cold regions of the shell, not to the shell as a whole. Likewise, a decrease in electron density ($n_e$) during the transition associated with recombination does not necessarily imply a corresponding reduction in the total gas density.

These comparisons show that the external ionization structure varies non-monotonically along the assumed star formation history (SFH). These variations are driven by the evolution of the radiation field, the properties of the superwind, and the gas distribution -- all of which are accounted for in the model sequences -- and cannot be explained solely by halo mass, age, or redshift. In particular, the non-monotonic behavior of the quantities $R_1$--$R_3$ across different sequences should not be interpreted as fluctuations of a single, continuous bubble. The sequences terminate and restart according to the stopping criteria described above; specifically, once Component 4 is absorbed by Component 3, the model switches to a different halo of comparable mass. The stellar population parameters continue to evolve in accordance with the prescribed star formation history.

In general, ionization profiles provide a physical interpretation of the contrasting behavior of Lyc and X-ray emission in existing models. Neutral regions within the shell and the surrounding gas correspond to states of Lyc suppression, whereas a predominantly ionized external medium is associated with significant Lyc escape. This correspondence is qualitative: the escaping flux depends on the integrated, frequency-dependent opacity, and a dominant ionized fraction does not, by itself, guarantee optical thinness. Together with the hot cavity, variations in the external ionization structure thus link the hydrodynamic evolution to both the escaping stellar continuum and the nebular contribution to the emergent spectrum.

\subsection{Evolution of the emergent spectra}

\begin{figure*}
\centering
\includegraphics[width=1.00\textwidth]{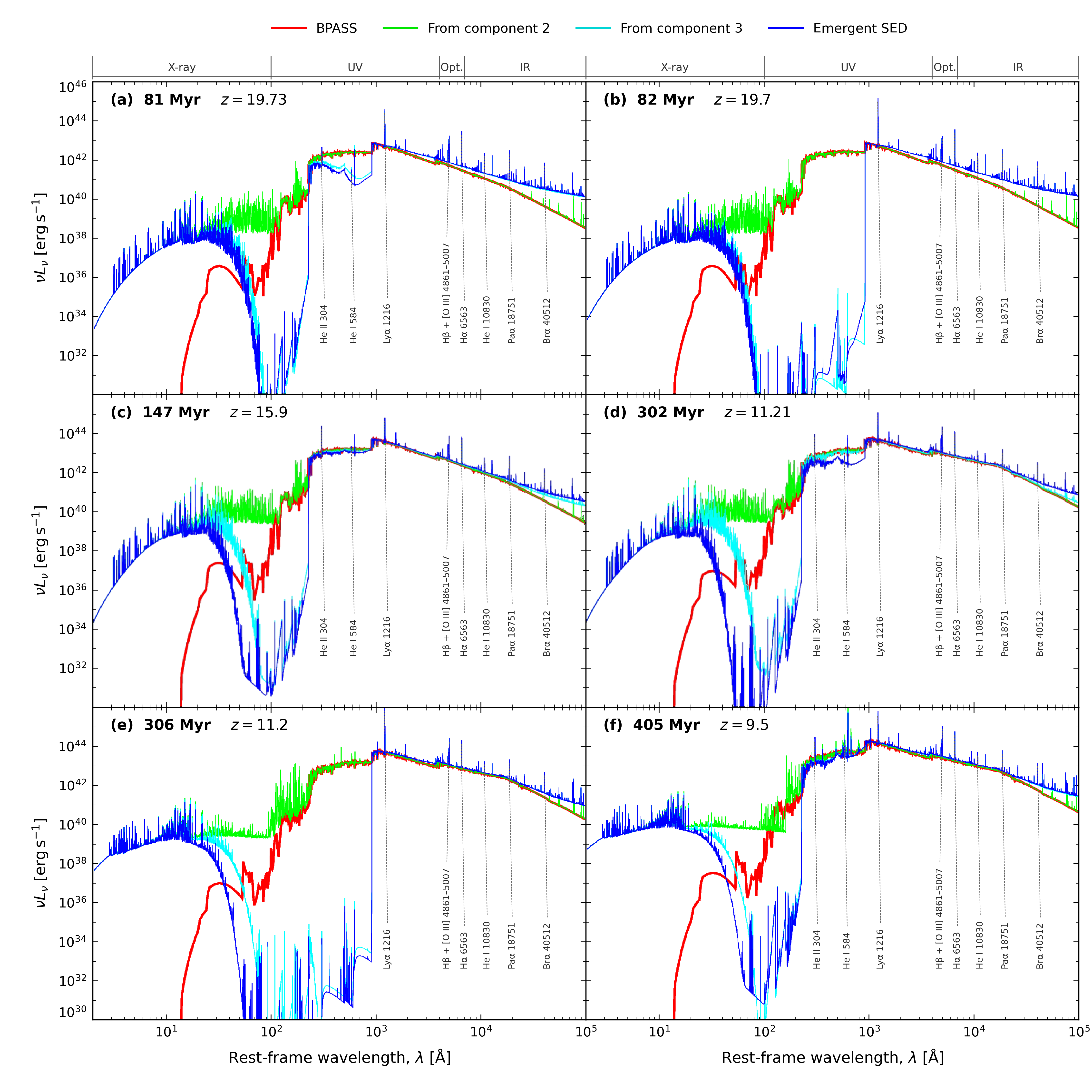}
\caption{
The evolution at the ages of 81\textendash{}405~Myr of the 2779 galaxy SFH derived in 
\citet{Brummel-Smith-2023} of the stellar spectrum\textendash calculated using the 
BPASS evolutionary stellar population synthesis method\textendash{}as radiation propagates 
through the components of a superwind bubble and the hydrodynamically undisturbed ambient 
medium of the nebular environment. The labels are the same as in Fig.~\ref{SEDs524}.
}
\label{fig:SEDs81405}
\end{figure*}

A comparative analysis of the modeled emergent spectra (see Figure \ref{fig:SEDs81405}) reveals substantial changes in both luminosity and spectral shape along the adopted formation history of galaxy 2779. The selected models span ages of 5--405~Myr, corresponding to a decrease in redshift from $z=28.73$ to $z=9.49$ and an increase in halo mass from approximately $1.7\times10^{7}$ to $2.4\times10^{10}\ M_\odot$. Here, age denotes the elapsed time along the adopted star formation history, rather than the age of a single stellar population. The spectra considered below are evaluated at the outer baryonic boundary, $R_4$, and therefore describe the radiation leaving the modeled galaxy before its subsequent propagation through the surrounding circumgalactic and intergalactic media.

A broad evolutionary trend is the increase in the luminosity of the emergent non-ionizing UV, optical, and infrared continuum. The non-ionizing UV continuum rises by several orders of magnitude between the earliest and latest models. This increase accompanies the growth of the stellar population and the evolution of the star formation rate and mechanical energy input. It should therefore be interpreted as the development of the composite stellar and gaseous system along the prescribed star formation history. Since halo mass, age, redshift, and source properties vary together in this sequence, their individual effects cannot be isolated from these models alone.

The most pronounced departures from a smooth luminosity evolution occur in the Lyman continuum. At 5 and 24 Myr, the emergent spectrum is strongly suppressed shortward of the hydrogen ionization edge. An appreciable escape of Lyc is present at 81 Myr, followed by a substantial suppression at 82 Myr. A comparatively transparent state is reached at 147 Myr, whereas the models at 302 and 306 Myr exhibit another transition from partial Lyc escape to strong attenuation. The final model, at 405 Myr, again produces a substantial emergent ionizing continuum. Thus, increasing halo mass does not lead to a monotonic increase in the transparency of the nebular medium.

The closely spaced models provide particularly useful comparisons. Between 81 and 82 Myr, the halo mass increases by only approximately 3.5~\%; between 302 and 306 Myr, the increase is approximately 2.4 percent. Nevertheless, the emergent Lyc changes markedly in both pairs. These differences demonstrate that the instantaneous structure and ionization state of the nebular medium play a major role in regulating the escaping radiation. Establishing the specific cause of each transition requires examination of the density distributions, ionic column densities, and dynamical states of the corresponding model components.

The X-ray spectrum also undergoes substantial evolution. The earliest model exhibits a weak, predominantly smooth X-ray hump, whereas a conspicuous emission-line spectrum is already present at 24 Myr. At later stages, the X-ray emission becomes more prominent relative to the non-ionizing UV continuum of the emergent spectrum. This development is accompanied by changes in the relative strengths of individual spectral features. The overall increase is not monotonic: the X-ray part of the spectrum at 302 Myr is weaker than at 147 Myr, followed by renewed enhancement at 306 Myr.

A preliminary identification based on the positions of the most conspicuous peaks suggests contributions from O VIII Ly-$\alpha$ at approximately 18.97~\AA, the O VII triplet at 21.6-22.1~\AA, Fe XVII transitions near 15.0 and 17.1~\AA, and C VI Ly-$\alpha$ at 33.74~\AA. The structured emission between approximately 12 and 14~$\AA$ is conservatively labeled as an unresolved Ne/Fe complex. The oxygen features are particularly prominent in several models between 24 and 302~Myr, whereas the peaks near 15 and 17~$\AA$ become especially conspicuous at 306 and 405~Myr. These identifications remain tentative pending verification against the model emission-line lists. Moreover, the peak heights in the plotted $\nu L_\nu$ spectra do not directly measure integrated line luminosities.

The evolution of the emergent X-ray spectrum reflects both changes in the emitting gas and wavelength-dependent attenuation by the surrounding components. In the later models, the long-wavelength side of the emergent X-ray hump is more strongly suppressed relative to the intermediate spectra. Consequently, changes in the relative prominence of oxygen and iron features cannot be attributed uniquely to a change in temperature. Their interpretation also requires the ionic abundances, elemental abundances, and intervening optical depths. The planned analysis of the radial temperature, density, and ionization distributions will therefore provide an important complement to the spectral comparison.

The Lyc and X-ray emission exhibit distinct responses to the evolving nebular structure. The X-ray spectra at 81 and 82 Myr remain similar despite the large change in Lyc escape. Likewise, the enhanced X-ray emission at 306 Myr coexists with a strongly attenuated Lyc. A single description in terms of progressive spectral hardening is therefore insufficient: different wavelength ranges respond differently to the changing physical conditions.

At non-ionizing wavelengths, the overall continuum shape evolves more gradually, while prominent emission features and an extended infrared continuum remain present. The available spectra establish the persistence of these features, but do not by themselves demonstrate a monotonic evolution of their equivalent widths or continuum slopes. Quantitative measurements require the numerical spectra and a consistent treatment of spectral resolution. Furthermore, because the selected models deliberately sample contrasting states and extrema, they do not determine the duration or relative frequency of the different spectral regimes.

Several features predicted by the models have direct counterparts in observations of early low-mass galaxies. In MACS0647-JD at $z=10.17$, \citet{Hsiao2024MIRI} detected H$\alpha$ and the [O III] doublet with JWST/MIRI. These are the same transitions identified in the optical part of our model spectra. NIRSpec observations of this system additionally revealed C~III], [O~II], [Ne III], H$\delta$, H$\gamma$, and the auroral [O III] $\lambda4363$ line \citep{Hsiao2024NIRSpec}. These measurements provide a useful set of diagnostics for testing the excitation and thermal conditions predicted by the models.

A complementary example is the Firefly Sparkle at $z=8.296$, a low-mass system with strong nebular lines and a substantial nebular continuum contribution in its central region \citep{Mowla2024}. Although observed at a lower redshift than the endpoint of our sequence, it illustrates the importance of including nebular emission when interpreting the spectra of young, low-mass galaxies. In particular, the emergent continuum need not provide a direct measure of the stellar contribution alone.

The comparison is more demanding for Ly$_\alpha$, whose propagation beyond the model boundary must be taken into account. Its detection in JADES-GS-z13-1-LA at $z\simeq 13$ demonstrates that Ly$_\alpha$ can emerge from very early systems under suitable conditions \citep{Witstok2025}. However, the model Ly$_\alpha$ luminosity at $R_4$ cannot be compared directly with an observed flux without accounting for external radiative transfer. The soft X-ray features discussed above remain predictions of the present models; we have not established observational detections of the corresponding individual lines in comparable Cosmic Dawn dwarf galaxies.

The observational comparisons therefore support the inclusion of nebular line and continuum emission in the interpretation of early galaxies, while a quantitative assessment of the present models requires comparisons of line fluxes, diagnostic ratios, equivalent widths, and continuum shapes at matched spectral resolution and aperture. Any revision of the stellar mass inferred for galaxy 2779 will require a dedicated fit to its own observations with the stellar and evolving nebular contributions treated consistently. Along the modeled formation history, the principal spectral result is a broad increase in luminosity accompanied by strongly non-monotonic Lyc escape and a changing X-ray line spectrum. These behaviors show that the emergent radiation retains information about the instantaneous state of the nebular medium in addition to the accumulated growth of the galaxy.

\begin{figure*}[t]
\centering
\begin{minipage}[t]{0.495\textwidth}
\vspace{0pt}
\includegraphics[width=\linewidth]{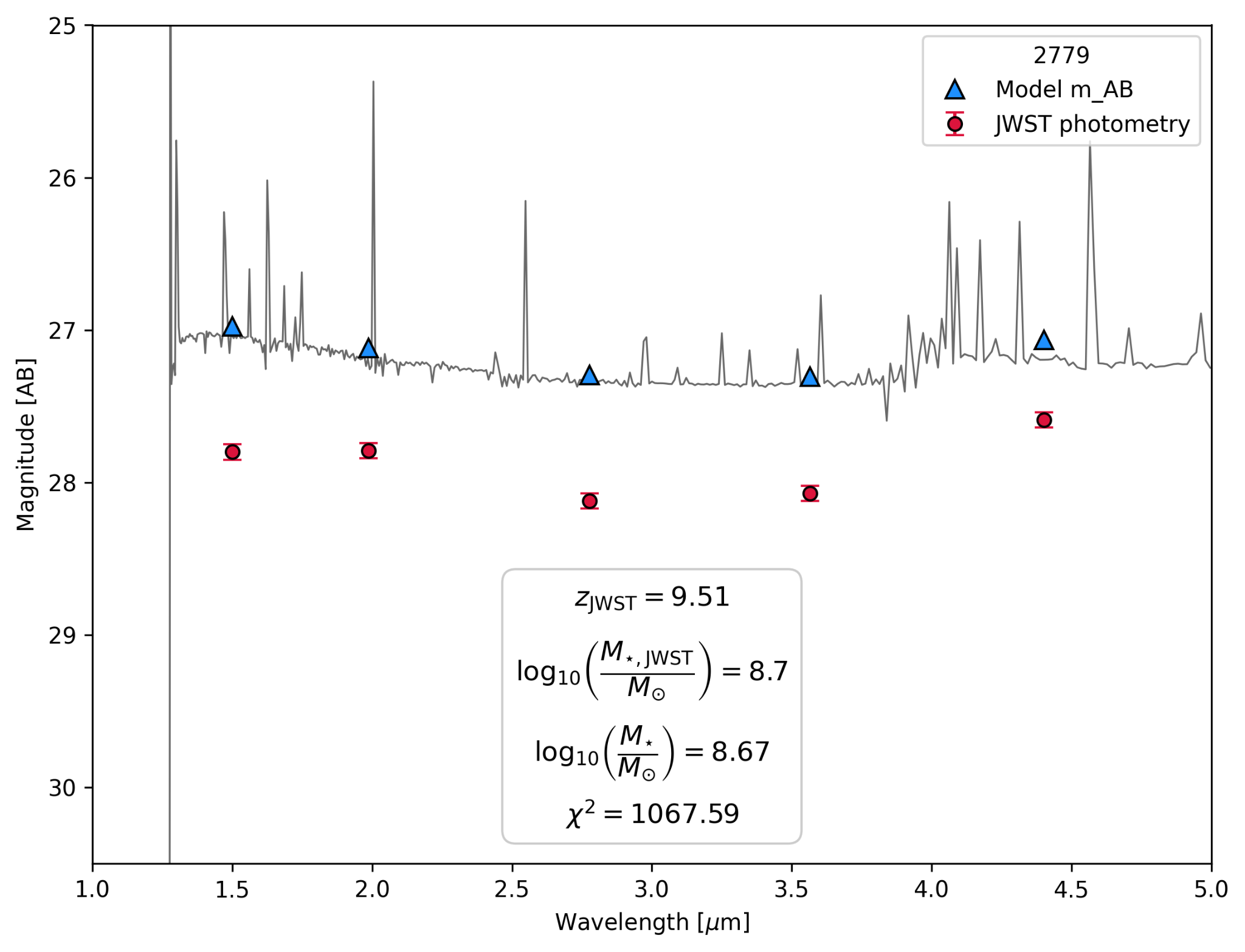}
\end{minipage}\hfill
\begin{minipage}[t]{0.495\textwidth}
\vspace{0pt}
\includegraphics[width=\linewidth]{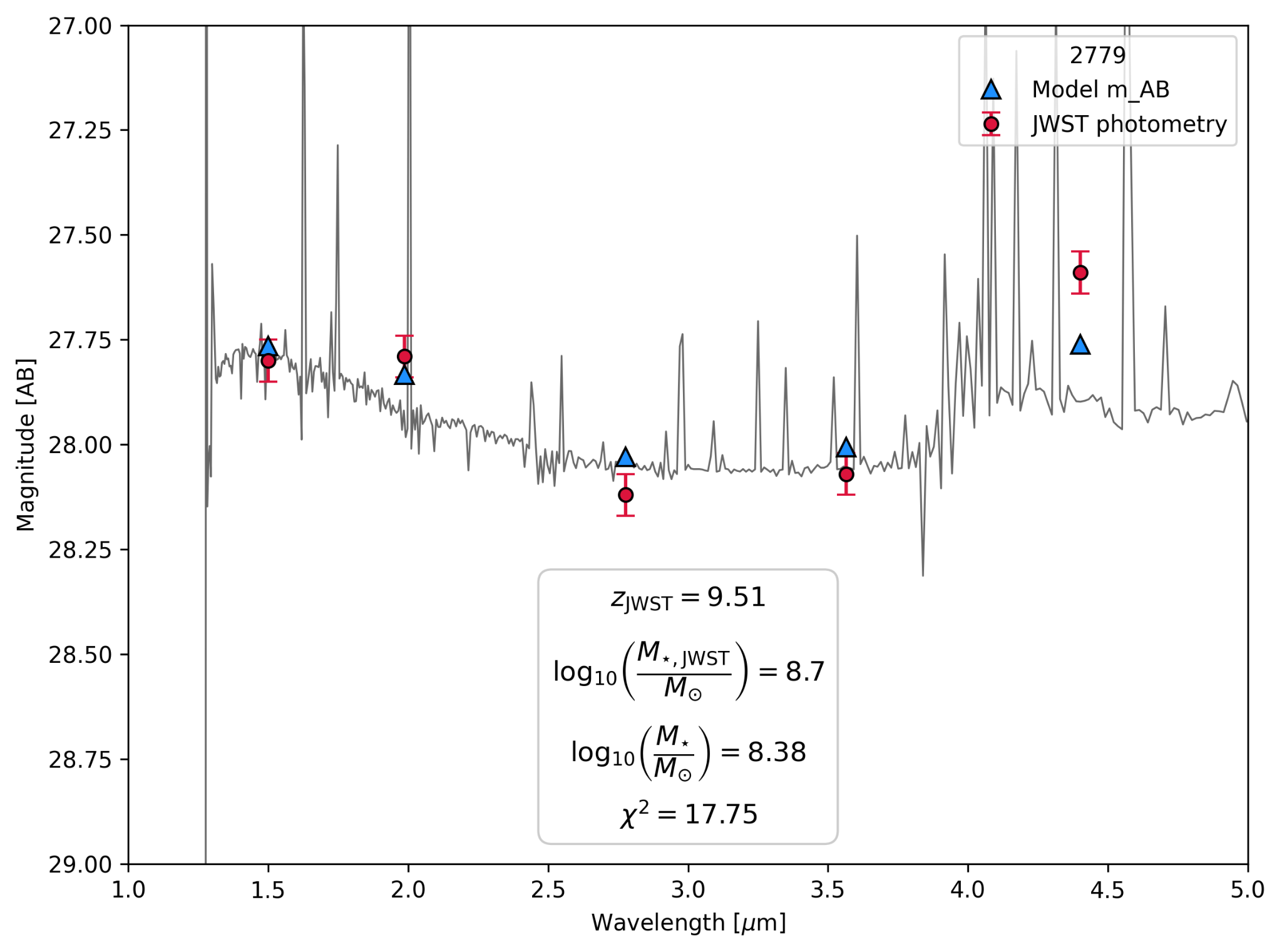}
\end{minipage}
\caption{Comparison of model SEDs of galaxy 2779 with the observed SED at z=9.51. While the red circles with error bars correspond to the
JWST photometry, the gray curves represent the MPhM SEDs in AB magnitudes, and the blue triangles show the model band-averaged
magnitudes in the following filters of NIRCam on board JWST: F150W, F200W, F277W, F356W and F444W. For more details see 
\protect\citet{Brummel-Smith-2023}.}
\label{fig:Mod_Obs}
\end{figure*}

\subsection{Comparison with JWST photometry}

\citet{Brummel-Smith-2023} compare the SEDs at the final stages of the reconstructed star formation histories of 12 dwarf galaxies (including galaxy 2779) with photometric data from the JWST. Our calculations, performed using the MPhM method, additionally account for radiative transfer within the nebular environment surrounding the star-forming region -- encompassing both the continuum emission and the line emission generated by the gas. Therefore, it is important to compare the spectrum resulting from the final stage of the reconstructed star formation history -- modeled using our approach for stellar and nebular emission -- with the observational data.

The left panel of Fig.~\ref{fig:Mod_Obs} compares the output spectrum of our final MPhM model with JWST photometric data for galaxy 2779. While the model generally reproduces the overall shape of the observed SED, it significantly overestimates the flux level. The left panel shows the model with a stellar mass of $\log (M_\star/M_\odot) = 8.67$ corresponding to the best-fit mass found by \citet{Brummel-Smith-2023}. The right panel compares the observations with a model having a stellar mass of $\log (M_\star/M_\odot) = 8.38$.

Accounting for the nebular contribution can alter the stellar mass required to reproduce the observational results. This possibility is illustrated in the right panel, which employs an earlier model from the evolutionary sequence -- one with a lower stellar mass -- rather than the final model. This comparison suggests that a lower stellar mass could improve the agreement with the photometric data. However, to derive a refined stellar mass, the SFH must be redetermined through a new model fit, with the output spectra generated by our approach consistently integrated into the fitting procedure.

This analysis is computationally intensive and is currently underway. We plan to present the results in a future publication. Applying this approach to galaxy 2779 also necessitates a corresponding re-analysis of other galaxies from the sample presented in \citet{Brummel-Smith-2023}, which further significantly increases the computational time required for this task.

Thus, the contribution of this work lies in the analysis of the evolving gaseous nebular environment and its relationship to the emission spectrum, as well as in demonstrating the possibility of improving the agreement between model spectral energy distributions (SEDs) and JWST photometric data by revising the estimated stellar mass.

\section{Conclusions}

The main conclusions of the present work can be summarized as follows.

(i) We constructed evolutionary sequences of MPhM (nebular photoionization) models for a redshift range from $z \approx 29.5$ to $z \approx 9.5$, based on the star formation history (SFH) determined for galaxy 2779 in \citet{Brummel-Smith-2023}. The calculations combine the evolution of the stellar population (modeled using the BPASS evolutionary stellar population synthesis method) with the evolution of the "superwind" components and the surrounding nebular medium, which remains hydrodynamically unperturbed.

(ii) The resulting spectrum includes stellar radiation that has passed through the medium, nebular emission lines, and the nebular continuum. Hot, diffuse gas within the cavity created by the superwind provides an additional contribution in the X-ray range.

(iii) At early stages, the shell can significantly suppress the escape of ionizing radiation (Lyc) while allowing X-rays to pass through. A noticeable escape of Lyc photons is observed at approximately $z \approx 20$ within the modeled evolutionary sequence.

(iv) The thermal and ionization structures provide a physical explanation for the differences in the escape regimes of ionizing radiation. As expected, in models with a strongly suppressed escape of Lyc photons, the outer part of the nebula consists primarily of neutral gas, and neutral regions are also present within the swept-up shell. In contrast, in models with a significant escape of Lyc photons, the outer gas is predominantly ionized. The hot, highly ionized gas within the superwind cavity provides an additional source of X-ray emission, which can emerge through the surrounding gas even when the escaping Lyc flux is strongly suppressed.

(v) We derived the dependencies of $f_{\rm esc}$ and $L_X/{\rm SFR}$ on time for the assumed star formation history. The nature of their variation -- specifically the increase during the final stage of the formation of galaxy 2779 -- reflects the coupled evolution of the stellar source and the surrounding gas. The rise in the star formation rate (SFR), as determined in the study by \citet{Brummel-Smith-2023}, leads to an increase in the production rate of stellar ionizing photons, the mass inflow rate into the superwind, and the superwind's mechanical luminosity.Together with the baryonic gas distribution, these evolving inputs determine the bubble structure, radiative transfer, and nebular line and continuum emission, including diffuse ionizing radiation. Due to the additional contribution from the nebular medium -- disturbed by the expansion of the superwind --the value of $f_{\rm esc}$ can exceed unity ($f_{\rm esc} > 1$). This is based on the normalization we have adopted, in which the total number of ionizing photons escaping the system per unit time, $Q_{\rm esc}$, is divided by the corresponding rate at which ionizing photons are generated by stars, $Q_*$: $f_{\rm esc}=Q_{\rm esc}/Q_*$. Thus, this quantity also accounts for ionizing photons generated by the gas itself and should be distinguished from the escape probability of purely stellar photons. This is a preliminary result. Whether the revised stellar mass, along with a corresponding self-consistent recalculation, will bring this ratio back into the range $(0 \leq f_{\rm esc} \leq 1)$ remains to be determined through a re-fitting of the parameters.

(vi) Individual model sequences terminate when a stagnation criterion is met or when component 4 is completely absorbed by component 3. Consequently, subsequent evolution is not accounted for in the calculated radiative contribution. 

Once either of the above stopping criteria for the modeled bubble expansion is met, we continue the evolutionary sequence by considering the formation and expansion of a new bubble, while retaining the prescribed evolution of the stellar population synthesis parameters that provide its stellar input. In the case of stagnation, we assume that the previous superwind bubble undergoes rapid collapse before the new bubble forms. When component 4 has been completely swept into component 3, we instead consider another halo of comparable mass at the corresponding evolutionary stage. We assume that the combined effects of halo growth and internal star formation have produced a new bubble in this halo, without explicitly modeling the coupling between these processes.

If this assumption is applied within a cosmological context, then calculations of this type, performed over a larger sample of galaxies, will provide an estimate based on a conservative criterion regarding whether dwarf galaxies can independently drive the reionization of the Universe. If the resulting predictions for $f_{\rm esc}$ and $L_X/{\rm SFR}$ prove sufficient to reproduce the required reionization history -- even without accounting for radiation emitted during the omitted evolutionary stages -- this would confirm the hypothesis that dwarf galaxies are sufficient within the adopted model framework. For this conclusion to hold, it is not strictly necessary that the specific parameter values represent absolute lower bounds.

(vi) Applying the MPhM method to cosmological estimates of $f_{\rm esc}$ (the ionizing radiation escape fraction) and the $L_X/{\rm SFR}$ ratio (X-ray luminosity to star formation rate) requires constructing a parameter grid that covers the ranges of halo masses, redshifts, and star formation histories (SFH) necessary to generate a representative population of galaxies. Such a grid can be created using cosmological simulations, for instance, IllustrisTNG50-1 \citet{nelson2019,pillepich2019}. We address this approach in a parallel study.

(vii) In \citet{Melekh_2024}, we demonstrated the importance of detailed modeling of diffuse ionizing radiation transport within complex geometries. Quantifying the errors introduced by the on-the-spot approximation in simulations such as SPHINX requires controlled comparisons with otherwise equivalent calculations that explicitly transport diffuse ionizing radiation. We plan to undertake such comparisons in future work by extending the current spherically symmetric modeling framework -- in which the outward-only approximation often provides a satisfactory treatment \citep{Buhajenko-Melekh-2016} -- to 2D and 3D geometries. Also this development will involve replacing classical self-similar solutions with magnetohydrodynamic simulations using codes such as \textsc{pluto} \citep{Pluto}, and adapting the MPhM approach developed by \citet{Melekh_2024}, which includes detailed transport of diffuse ionizing radiation, to the first dwarf galaxies.

This work establishes a foundation for a program aimed at progressively increasing the complexity and detail of modeling the spectral energy distributions (SEDs) of dwarf galaxies during the Cosmic Dawn. Such a foundation is essential for evaluating the discrepancies between simplified and more sophisticated models, particularly regarding their predicted ionization structures and the escape fractions of ionizing photons. Furthermore, it will enable us to investigate how these differences influence the outcomes of cosmological simulations concerning the neutral hydrogen signal (at the 21 cm wavelength), as well as the spectral characteristics of the first molecules that formed during the Dark Ages and the onset of the Cosmic Dawn.

\begin{acknowledgements}

This work is done in the framework of the project “Tomography of the Dark Ages and Cosmic Dawn in the lines of
hydrogen and the first molecules as a test of cosmological models” (state registration number 0124U004029) supported
by the National Research Foundation of Ukraine.

\end{acknowledgements}

\bibliographystyle{aa}
\bibliography{references}

\end{document}